\documentclass{aastex}          
\usepackage{spr-astr-addons}    

\usepackage[hyperindex,breaklinks]{hyperref}

\newcommand{\kms}{\hbox{km$\,$s$^{-1}$}}

\newcommand{\gf}{\hbox{\it gf}}
\newcommand{\CMFGEN}{\hbox{\sc cmfgen}}

\begin{document}
%
\title{The Atomic Physics Underlying the Spectroscopic Analysis of Massive Stars and Supernovae}

\shorttitle{Spectroscopic Analysis of Massive Stars and Supernovae}
\shortauthors{D. John Hillier}

\author{D. John Hillier \altaffilmark{1}} 
\affil{Department of Physics and Astronomy, \\
3941 OÕHara Street,\\ 
University of Pittsburgh, Pittsburgh, PA, 15260}
\email{hillier@pitt.edu} 


\begin{abstract}
We have developed a radiative transfer code, \CMFGEN, which allows us to model the spectra of massive stars and supernovae.  Using \CMFGEN\ we can derive fundamental parameters such as effective temperatures and surface gravities, derive abundances, and place constraints on stellar wind properties. The last of these is important since all massive stars are losing mass via a stellar wind that is driven from the star by radiation pressure, and this mass loss can substantially influence the spectral appearance and evolution of the star.  Recently we have extended \CMFGEN\ to allow us to undertake time-dependent radiative transfer calculations of supernovae. Such calculations will be used to place constraints on the supernova progenitor, to place constraints on the supernova explosion and nucleosynthesis, and to derive distances using a physical approach called the ``Expanding Photosphere Method''. We describe the assumptions underlying the code and the atomic processes involved.  A crucial ingredient in the code is the atomic data. For the modeling we require accurate transition wavelengths, oscillator strengths, photoionization cross-sections, collision strengths, autoionization rates, and charge exchange rates for virtually all species up to, and including, cobalt. Presently, the available atomic data varies substantially in both quantity and quality.
\end{abstract}


%
\section{Introduction}

Massive stars are a crucial ingredient of galaxies, and the universe. They enrich the interstellar medium (ISM) with metals, either through ``quasi-steady'' mass loss, or when they explode as supernovae (SNe). They deposit momentum and energy into the ISM, and ionize the surrounding gas, producing colorful nebulae (H\,{\sc ii}  regions). Massive stars also exhibit a wide range of interesting phenomena  including radiation driven winds, and colliding winds. Colliding winds, which occur in massive star binaries,  generate strong shocks that give rise to hard X-ray emission \citep[see, e.g., review by][]{Ste05_xray}. The key tool for understanding massive stars is spectroscopic analysis.


Using spectroscopic analysis we generally wish to infer the fundamental parameters that describe the star --- its mass, effective temperature, radius, surface gravity, and surface abundances. From these, and with the aid of evolutionary models, we try to infer both the previous evolutionary history of the star and its future evolution. As we generally can't study the evolution of a single star we need to infer/constrain the effects of complex physical processes by studying groups of stars.  Constraints provided by these studies can then be used to improve evolutionary calculations.

There are still significant uncertainties in evolutionary models (e.g., treatment of convection, stellar rotation, and time-dependent mass loss). Rotation, for example, is now recognized as a crucial factor in massive star evolution --- it affects the evolutionary lifetime of the star, surface abundances, and the star's evolution, and there is a strong interaction between rotation and mass loss  \cite[e.g.,][]{MM00_evol_rot}. Since stars form with a range of rotation rates, it is no longer possible to assign a position in the HR-diagram to a unique mass \cite[e.g.,][]{MM00_evol_rot}.

There are also uncertainties in spectroscopic analyses. Some of  these arise from model assumptions  and inadequacies in the atomic data. Others arise from an incomplete understanding of the stars we are trying to model. O stars, for example, show strong evidence for microturbulent velocities approaching the sound speed, while Of stars show evidence for macroturbulent velocities in excess of the sound speed. The origin of these velocity fields is unknown. Mass-loss rates are a key ingredient of stellar evolution models, but deriving accurate mass-loss rates is difficult. While wind theory provides a qualitative understanding of mass loss in O stars, there are fundamental uncertainties since theory predicts, and observations show, that the winds are highly inhomogeneous.  As a consequence of the inhomogeneities, mass-loss rates derived from observation depend on the diagnostic used. For more information on radiation driven winds, mass-loss rates, and problems modeling O stars, the reader is referred to reviews by \cite{2008IAUS..250...89H, 2008A&ARv..16..209P} and  \cite{2009AIPC.1171..173O}.

\section{CFMGEN --- A Spectroscopic Tool}

In order to facilitate analysis of hot stars (stars in which molecules and energy transport by convection at the stellar surface can be neglected) we have developed a non-LTE radiative transfer code, \CMFGEN\
(Hillier \citeyear{Hil87_A, 1990A&A...231..116H};  Hillier \& Miller \citeyear{HM98_blank}).
 
The primary purposes of \CMFGEN\ are as follows:
 \begin{enumerate}
 \item
To derive accurate stellar parameters and abundances for comparison with ``evolution'' calculations. 
 \item
To provide accurate EUV (i.e., $\lambda <  912$\,\AA) radiation fields for input to nebular photoionization calculations. 
\item
To provide fundamental data for the study of starbursts, star formation in galaxies, etc. 
\item
To provide a better understanding of the hydrodynamics of stellar winds. 
\item
To provide distances to Type II SNe using the expanding photosphere method \citep[EPM --- e.g.,][]{ESK96_EPM, BNB04_1999em, DH06_SN1999em} and its variants. 
\item
To provide diagnostics of SNe which can place constraints on the progenitor and the explosion.
\item
To allow the development and testing of  approximate methods that can be used in more complex geometries and in inhomogeneous media. 
\end{enumerate}

\noindent
{\sc cmfgen} has been used to study O  Stars; Wolf-Rayet (W-R) Stars; Luminous Blue Variables (LBVs);
A \& B supergiants;  and  Type  I and Type II Supernovae.

{\sc cmfgen} was originally designed to model spectra of W-R stars which have a dense stellar wind. The hot W-R core, together with the dense stellar wind, gives rise to an optical spectrum dominated by emission lines --- the antipathy of normal spectra in which the optical spectral region is dominated by absorption lines. Because of the stellar winds, with flow velocities of order 1000\,\kms, it is convenient to solve the radiative transfer equation in the comoving frame --- in this frame the opacities and emissivities can be assumed to be isotropic.

 In \CMFGEN, the primary radiation transport equations to be solved (assuming $V \propto r$ and spherical geometry) are

\begin{equation}
  {1 \over cr^3}  {D(r^3 J{_\nu})  \over Dt} + {1 \over r^2} {\partial (r^2 H_\nu)  \over \partial r}
  - {\nu V \over rc} { \partial J_\nu \over \partial \nu } = \eta_\nu - \chi_\nu J_\nu
 \label{eq_zero_mom}
 \end{equation}
\noindent
and

\begin{eqnarray}
  {1 \over cr^3}  {D(r^3 H_\nu)  \over Dt} + {1 \over r^2} { \partial(r^2 K_\nu)  \over \partial r}
 \,\,  + && \hskip -0.65cm {K_\nu - J_\nu \over r} \nonumber \\
  - && \hskip -0.65cm {\nu V \over rc}{ \partial H_\nu \over \partial \nu } = - \chi_\nu H_\nu 
\label{eq_first_mom} \,\,.
 \end{eqnarray}

\noindent
In the above 

\begin{equation}
[J_\nu,H_\nu,K_\nu]= {1 \over 2} \int^{1}_{-1} [1,\mu,\mu^2] I_\nu(t,r,\mu) d\mu
\end{equation}

\noindent
are moments of the radiation field, $D/Dt$ is the Lagrangian derivative, $I_\nu$ is the specific intensity at time $t$, location $r$, frequency $\nu$, and in direction $\mu$,  $\mu=\cos \theta$ where $\theta$ is the angle between the radius vector at $r$ and the specific intensity, $\chi_\nu$ is the (frequency dependent) opacity, and $\eta_\nu$ is the emissivity \citep[e.g.,][]{Mih78_book, MM84_FRH_book}.  When written in this form, the equations are deceptively simple --- in reality they are very complex. First, these equations need to be solved at a large number of frequencies, typically 100,000. Due to the $\partial/\partial \nu$ term, these equations are explicitly coupled in frequency. Second, much of  the physics is hidden in $\chi_\nu$ and $\eta_\nu$. In the best case scenario these are determined by the local temperature and density, but even then the temperature is determined by the radiation transport, and thus there is an implicit coupling of the opacities and emissivities with the radiation field.  

In stellar atmosphere modeling we can neglect the Lagrangian derivative term, and
the equations are easily solved (for non-Hubble flows additional terms are introduced).
However, for Type II SNe, it has become increasingly apparent that to accurately model the spectra the full time-dependent radiation transport equations must be solved.

\section{LTE \& non-LTE}

A great simplification, applicable to most main-sequence spectral types, is the assumption of {\it local thermodynamic equilibrium (LTE)}. With this assumption we assume that the state  of the gas is entirely determined by its temperature and density (and composition). Thus we can use statistical arguments to determine the ionization state of the gas (Saha equation) and level populations (Boltzmann or Saha-Boltzmann equation). The LTE assumption is valid when collisional processes (which couple directly with the local gas) dominate over radiative processes (which couple the gas to gas elsewhere by radiation). 

In LTE, we require line lists and opacities (due to bound-free, free-free, and bound-bound processes). The opacities, for a given composition, are simply functions of the density and electron temperature. Deep inside the star radiation diffuses, and the transport of radiation is dictated by the Rosseland mean opacity defined by

\begin{equation}
{1\over \chi_R} = {\pi \over 3 \sigma T^3} \int_0^\infty {1\over \chi_\nu} {dB_\nu(T) \over dT} \,d\nu \; .
\end{equation}

\begin{figure*}
\includegraphics[scale=1.0,angle=-90]{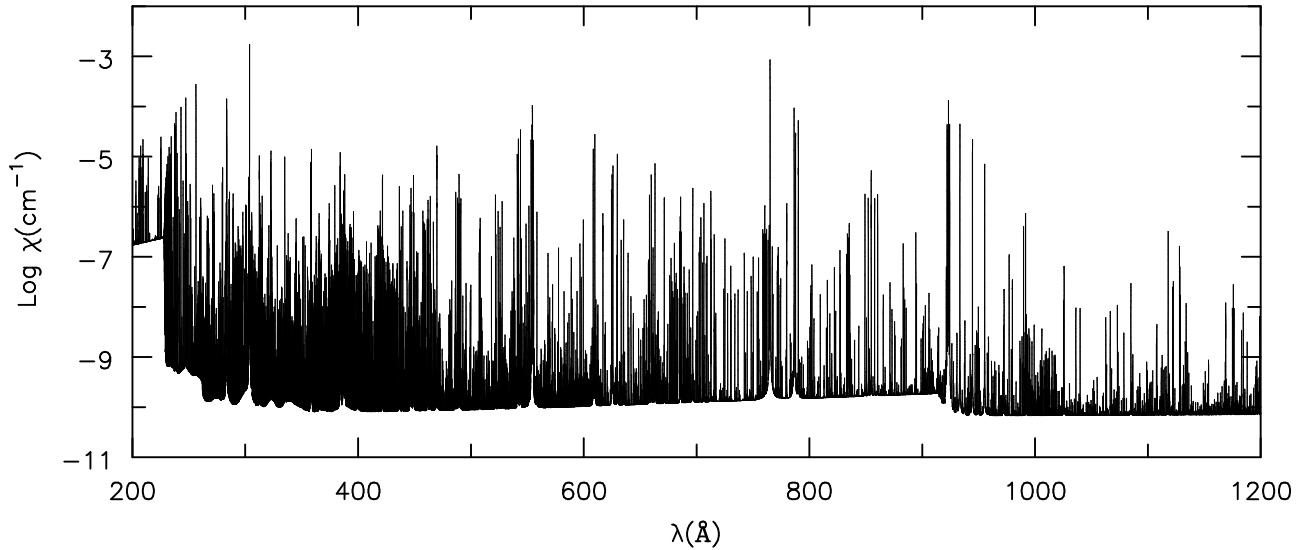}
\caption[]{Illustration of the complex opacity at one location (with $T \sim 40,000\,$K, $N_e \sim 10^{15}$\,cm$^{-3}$) for a typical
model atmosphere.}
\label{Fig_opac}
\end{figure*}

In the photosphere the situation is very different --- use of the Rosseland mean opacity is no longer applicable and we must solve the radiative transfer equation at every frequency. To do this we need a detailed description of the opacities (Fig.~\ref{Fig_opac}). Since the theoretical spectrum is to be compared with observation, it is important the bound-bound transitions have accurate wavelengths.

In O stars, W-R stars and SNe (as well as many other objects) the assumption of LTE is invalid. Instead we are forced to solve the equations of statistical equilibrium --- the equations describing how individual levels in an atom are populated and depopulated. Because the radiation field is no longer Planckian (and described by the local electron temperature) the atomic populations that satisfy these equations will generally differ from their LTE values. In order to solve for the populations many atomic processes need to be treated. These include:
\begin{enumerate}
\item
Photoionization and radiative recombination (i.e., bound-free processes)
\item
Low and high temperature dielectronic recombination (LTDR and HTDR). In high temperature (or classical) dielectronic recombination, recombination occurs through doubly excited autoionizing  states with large {\it n} \citep[e.g.,][]{Bur64_HTDR}. For example, the 2p\,nl states in C\,{\sc iii} that converge on the C\,{\sc iv} 2p state, and which ``recombine''  through the decay of the 2p electron.  In contrast, LTDR usually refers to recombination through doubly excited states that lie close to, but above the ionization limit \citep{NS83_LTDR}.  As a consequence, LTDR rates are sensitive to the atomic structure --- a single energy level close to the ionization edge can dominate the recombination rate at low temperatures
\item
Bound-bound transitions
\item
Collisional excitation and de-excitation. In hot stars this occurs primarily by electrons, but in cooler stars collisions with other species are also important. Fortunately, the conditions in stellar atmospheres are such that the electrons have a Maxwellian velocity distribution,
and thus we require collision strengths averaged over a Maxwellian velocity distribution. \item
Collisional ionization and collisional recombination (i.e., three-body recombination)
\item
Auger ionization by X-rays (and gamma-rays in SNe)
\item
Charge exchange reactions
\item
Two-photon emission
\item
In SNe, collisional ionization and excitation with non-thermal (high-energy) electrons that are created via
gamma-ray photons and Compton scattering \cite[see, e.g.,][]{KF92_gam_rays}.
\end{enumerate}

When non-LTE is applicable, there is a tremendous non-linear coupling between the radiation field and the level populations. The radiation field determines the electron temperature and level populations which, in turn, determine the radiation field.  As a consequence iterative techniques are used in order to obtain full consistency between the radiation field and level populations \citep[see review by][]{2009AIPC.1171....3H}.

In Type II SNe, even greater complexities are introduced. To accurately model Type II SN spectra  it  is important to include advection terms, which couple the populations at one time step to those at an earlier time step, into the equations of statistical equilibrium. 
The  inclusion of advection terms into the rate equations helps explain the strength of H$\alpha$ emission in spectra of SN1987A \citep{UC05_time_dep, DH10_tdrt} and SN1999em \citep{DH08_time}.

\section{Atomic Details Matter}
\label{Sec_at_mat}

In the construction of model atmospheres the precise details of the atomic opacity are generally unimportant. However in spectroscopic analyses this is not the case. In hot stars we often have only a few lines that arise from a given species/ionization stage. In order to deduce abundances, it is crucial that our model fully describe the formation process of each individual line. In LTE this is  relatively easy. In non-LTE this can be very difficult, since we need to fully understand the processes affecting the populations of the levels involved in the transition. This means having all the atomic data relevant to those levels and, since the population of those levels (except in a few special circumstances) depends on the population of other levels in the atom, we  require atomic data for the whole atom.  Unfortunately the situation can be even more complex than this --- in some cases line strengths of one species are affected by complex interactions with another species. To illustrate the complexities, we discuss four examples.

In WC stars, evolved massive stars deficient in H, and showing He, C, and O emission lines in their spectra, C\,{\sc iii} $\lambda$5696 is used as one of the key classification diagnostics. In WC4 stars the line is very weak, if not absent, while its strength increases (but with scatter) as we move along the spectral sequence from WC4 to WC8 \citep{TCM86_WC_stars}. Interestingly, other C\,{\sc iii} lines show much less variation in strength along the spectral sequence \citep[e.g.,][]{1990ApJ...354..359C}; thus there is something special about the formation of C\,{\sc iii} $\lambda 5696$.

\begin{figure}
\includegraphics[scale=0.48]{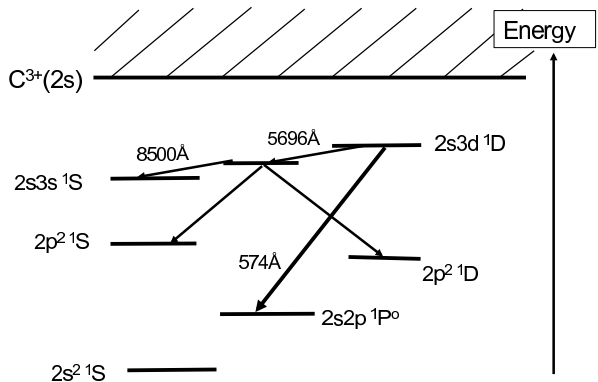}
\caption[]{Simplified Grotrian diagram (not to scale) for C\,{\sc iii}.}
\label{Fig_CIII_5696}
\end{figure}

A simplified Grotrian diagram for the singlet-terms of C\,{\sc iii} is shown in Fig.~\ref{Fig_CIII_5696}. C\,{\sc iii} $\lambda 5696$ emission is produced by the decay of the 2s\,3d\,$^1$D state to the 2s\,3p\,$^1$P$^{\hbox{o}}$ state ($A=4.3 \times 10^7$\,s$^{-1}$). However, photons prefer to decay from the 2s\,3d\,$^1$D to the 2s\,2p\,$^1$P$^{\hbox{o}}$ state, with $A=6.3 \times 10^9\,$s$^{-1}$. In early WC stars, most of the decays occur via this route --- it is only when this transition becomes optically thick, that C\,{\sc iii} $\lambda 5696$ is driven into emission \citep{H89_WC}. To complicate matters further, the strength of C\,{\sc iii} $\lambda 5696$ is also influenced by low-temperature dielectronic recombination. Interestingly the 2s\,3p\,$^1$P$^{\hbox{o}}$ state preferentially decays to 2p$^2$\,$^1$D rather than 2s\,3s\,$^1$S \citep{N71_Of_emiss, 1978PhDT.........9C}. Thus $\lambda 5696$ can be in emission, while $\lambda$8500 (produced by the decay of 2s\,3p\,$^1$P$^{\hbox{o}}$, the lower level of $\lambda 5696)$ can remain in absorption --- a phenomena seen in Of stars \citep{EW81_CIII}.

Another line formation mechanism is continuum fluorescence, which is best illustrated by an example. In C\,{\sc iv} there is a strong transition at 312\AA\ which connects the ground state, 2s\,$^2$S,  to the 3p\,$^2$P$^{\hbox{o}}$ state. As we are dealing with a ground state transition, it is usually optically thick, and hence photons typically scatter many times in this transition before escaping,  or\ before being destroyed. However the 3p\,$^2$P$^{\hbox{o}}$ level can also decay via a transition at $\lambda \lambda 5801, 5812$ to the 3s\,$^2$S state. The probability of this occurring is low ($\sim$~1/170 per scattering) but it does provide a means of converting far UV photons into optical emission \citep{1988ApJ...327..822H}. To get the strength of the observed optical emission correct we have to have a  good understanding of the C\,{\sc iv} model atom as well as accurately model the spectral region around 312\AA\ --- a region which cannot be directly observed in O and W-R stars, and which suffers heavy line blanketing by iron group elements. Continuum fluorescence is important in many astrophysical objects including W-R stars belonging to the nitrogen sequence (WN stars), LBVs, and quasars. 

Perhaps the most famous example of line overlap influencing line strengths is seen in nebula spectra.  In nebula, some O\,{\sc iii} lines are seen to be unusually strong --- much stronger than would be predicted by recombination theory. Moreover, line strengths in individual multiplets do not correspond to those observed in the laboratory. The explanation lies in the chance coincidence of an O\,{\sc iii} line ($\lambda 303.80$) with He\,{\sc ii} Ly$\alpha$ ($\lambda  303.78$).  As the He\,{\sc ii} Ly$\alpha$ transition is optically thick, photons emitted in the transition are trapped and scatter many times. During this scattering process there is a chance that some of the Ly$\alpha$ photons will be absorbed by O\,{\sc iii}. The upper O\,{\sc iii} levels have several alternative decay routes, some of which lead to the enhanced emission in some lines seen at optical and UV wavelengths. The process is termed Bowen Resonance Fluorescence, after its discoverer \citep{Bow34_fluro}. A more recent and detailed discussion of the mechanism is provided by \cite{Ost89_book}.

In O stars a similar overlap occurs. In this case there is a chance overlap of Fe\,{\sc iv} lines with the He\,{\sc i} resonance transition at 584.33\AA. The Fe\,{\sc iv} lines remove photons from the transition, lowering the population of the  1s\,2p\,$^1$P$^{\hbox{o}}$ level, which in turn affects the strength of He\,{\sc i} singlet levels in the optical spectra \citep{NHP06_HeI}. This in turn affects effective temperature determinations, since the ratio of He\,{\sc i} to He\,{\sc ii} line strengths is used as a temperature diagnostic. Prior to the discovery of this effect it was known that there were inconsistencies between singlet and triplet He\,{\sc i} lines in some O stars. The effect was erroneously ascribed (at least in this author's opinion) to a problem with the triplet line strengths. However, in hindsight it is much more likely that the problem lies with the singlet line strengths because the lower level of most of the optical diagnostic lines is coupled to the  1s\,2p\,$^1$P$^{\hbox{o}}$ state whose population is affected by radiation-transfer effects in the resonance transition.

\section{Atomic Data Requirements}
Astrophysicists are interested in understanding the physical processes and properties
of astrophysical objects (stars, SNe etc). Ideally we would have all the required atomic data,
and any discrepancies between models and observations would only be related to model assumptions and the neglect of  crucial physical processes. Unfortunately this is not the case --- in the real world only limited atomic data is available, and it is of mixed quality. While great strides have been made improving the quality and quantity of atomic data (e.g., \citealt{Sea87_OP, HBE93_IP, Kur09_ATD}\footnote{Atomic data from Robert Kurucz is available at 
http://kurucz.harvard.edu/}), the availability of atomic data and its quality must always be considered when performing spectroscopic modeling to obtain results of astrophysical importance. Given limited resources, what are the most crucial data sets that are still required?

In general, the most important elements for hot stars are H, He, C, N, O (generally referred to as CNO elements), and Fe, with Ne, Si, S, and Ar of somewhat lesser importance. In SNe the situation is somewhat different, with other iron group elements (particularly Ni and Co) also being of crucial importance. Even for this small subset of species, important atomic data is missing.

One of the most important requirements, but perhaps the least appreciated, are accurate energy levels and wavelengths. In general, energy levels and wavelengths of sufficient accuracy can generally only be obtained from observation. As once noted to me, \gf\ values accurate to 10\% are great but a wavelength accurate to 10\% is (almost) useless. For CNO elements  atomic energy levels and wavelengths are generally pretty good, however additional data are still needed. This is especially true of the infrared spectral region. Additionally, accurate energy levels are needed for doubly excited states that lie near, or above, the ionization limit (e.g. C\,{\sc iii}) since the precise location of these states is important for determining dielectronic recombination rates, particularly at low temperatures. Measuring the width of lines from these states can also give an estimate of autoionization probabilities, providing an important check on theoretical calculations.

As regards \gf\ values, there has been a proliferation of theoretical data which has greatly facilitated the advances in spectroscopic analyses. In many cases, the calculations assume LS coupling which is generally adequate for computing mean opacities and model atmosphere structures, but may not be so useful for performing non-LTE abundance studies. Theoretical calculations can often provide reliable \gf\ values, but for some cases the \gf\ values are not sufficiently accurate for analyses. As an example we consider the three most important  decays from the 2p\,3p\,$^3$D state in C\,{\sc iii}. These are \\
\indent
2s\,2p\,$^3$P$^{\hbox{o}}$ -2p\,3p\,$^3$D (0.00926,  $2.7 \times 10^8$\,s$^{-1}$,       369.4\AA),\\
\indent
2p\,3s\,$^3$P$^{\hbox{o}}$ -2p\,3p\,$^3$D (0.166,  $1.5 \times 10^7$\,s$^{-1}$,    6740.6\AA), and \\
\indent
2s\,3p\,$^3$P$^{\hbox{o}}$ -2p\,3p\,$^3$D (0.00286,  $4.6 \times 10^6$\,s$^{-1}$,   1577.4\AA) \\where the numbers in brackets are the oscillator strength, A value, and wavelength, respectively. Using \gf\ values computed by \cite{NS84_CNO_LTDR}, Hillier (\citeyear{Hil87_A}) found that the 1577\AA\ line in the WC star HD\,165763  was much stronger than would be predicted on the basis of the observed strength of the 6740.6\AA\ line. Improved calculations by Peter Storey (private communication) indeed showed that the \gf\ value for the 1577\AA\ line was too large, and his revised value (given above) gave much better agreement with observation. Unfortunately, his calculations also revealed that the actual value was quite sensitive to assumptions used in the calculations. 

Another example is the \gf\ values of the Fe\,{\sc iv} transitions which overlap the He\,{\sc i} resonance transition (see \S \ref{Sec_at_mat}).   The two transitions of most interest are \\
\indent
 Fe\,{\sc iv}  3d$^5$ $^4$F$_{9/2}$  -- 3d$^4$($^3$G)4p $^2$H$^{\hbox{o}}_{9/2}$ and \\
\indent
Fe\,{\sc iv} 3d$^5$ $^2$D3$_{5/2}$ -- 3d$^4$($^3$G)4p $^4$H$^{\hbox{o}}_{7/2}$.  \\
\noindent
 Bell and Kurucz (\citeyear{1995KurCD..23.....K}) give \gf\ values of 0.00349 and 0.0288 for the two transitions, while \cite{1995A&A...301..187B} give 0.00251 and 0.00264. In these cases measurements are clearly desirable to constrain the strength of these lines.

Photoionization cross sections are now available for atoms with even atomic numbers, up to and including Fe, primarily through the OPACITY and Iron Projects  \cite[e.g.,][]{Sea87_OP, HBE93_IP}. These cross-sections appear to show reasonable agreement with experiment when available. One possible deficiency in the cross-sections is the location of the resonances, which are theoretical rather than experimental. Erroneous locations can potentially affect recombination rates (particularly at low temperatures) and there is also the potential problem of interactions between resonances and spectral features. The importance of the latter effect is unknown. To expedite calculations in \CMFGEN\ we generally smooth the photoionization cross-sections (typically with a Gaussian of full-width 3000\,\kms). Originally the smoothing was hard-wired into the cross-sections, but for most cross-sections the smoothing  is now a control parameter. This allows us to easily test the influence of the smoothing. As computers become faster, and especially for 1D models, smoothing is no longer a necessity, but rather a computational tool.

For other elements photoionization cross-sections are unavailable. Of particular importance are elements like Co and Ni which have high abundances in SNe. At present in \CMFGEN\  relatively crude approximations are used [data for Ni\,{\sc ii} is available ---  \cite{NB01_NiII}]. Lines belonging to Sc\,{\sc ii} can also be readily identified in Type II SNe spectra, and photoionization data for Sc ions is also unavailable. The case of Sc is interesting --- Sc has relatively low abundance and thus it has very little influence on SNe spectra. However it does produce readily identifiable features, and matching such features does provide a quantitative test of the spectroscopic model. It also pleases observers (and some theorists), who are quick to  point out the absence of Sc features from models.

The lack of collisional data is probably the area of most concern. For low lying levels, collisional data is generally available, but for higher levels such data is usually lacking. In such cases approximate formulae, of unknown accuracy, are often used. The formulae often depend on the \gf\ value of the transition connecting the two levels, but as has been pointed out in the past  \cite[e.g.,][]{Mih78_book}, collisional rates for LS semi-forbidden or forbidden transitions can be as large as those for non-forbidden transitions. For the vast majority of applications we need cross-sections averaged over a Maxwellian velocity distribution, and it is best tabulated as a temperature dependent collisional strength. Despite recent improvements \citep[e.g.,][]{PB04_Hcol} there are still uncertainties with the hydrogen collision strengths.

Charge exchange reactions are of particular importance in non-LTE modeling. In many cases a charge exchange reaction with H, rather than the recombination rate, determines the ionization balance. An excellent example is O$^+$ + H $\rightleftharpoons$ O + H$^+$, which has a  rate coefficient of order $10^{-9}$\,cm$^{3}$\,s$^{-1}$. Many cross-sections for charge reaction rates with H and He have been computed; a convenient tabulation is that of \cite{KF96_chg}. One problem with tabulated cross-sections is that the charge exchange channels are not always provided. Such information is necessary as we go to high density, since we must include the reverse reactions in order to recover LTE.

Another concern is the paucity of data for charge exchange reactions with species other than H and He. This paucity is understandable --- H and He have the greatest effect since they are easily the most abundant elements in most astrophysical contexts. However, there are objects that are H and/or He deficient. In WC stars for example, H is absent, and He, C, and O have comparable abundances. In SNe a wide range of species are present, with C, O, Si,  Ne, Fe, Ni, Co exhibiting large mass fractions in some regions. Despite advances in spectral modeling of SNe \citep[e.g.,][]{BBH07_SNeII, KTN06_SN_MC, Kas09_SNII, DH10_tdrt}, there are still significant uncertainties, and it is not currently possible to identify a discrepancy between model and observation that might be due to a charge exchange reaction.

\acknowledgments
I would like to thank the organizers of the 2010 HEDLA conference for a superb meeting. ÒSupport for program HST-AR-11756.01-A was provided by NASA through a grant from the Space Telescope Science Institute, which is operated by the Association of Universities for Research in Astronomy, Inc., under NASA contract NAS 5-26555.Ó 
 


%

%

\end{document}